**From diffusive to ballistic transport in etched graphene constrictions and nanoribbons**

*Sowmya Somanchi[1], Bernat Terrés[1,2], Julian Peiro[1], Maximilian Staggenborg[1], Kenji Watanabe[3], Takashi Taniguchi[3], Bernd Beschoten[1], and Christoph Stampfer[1,2,\*]*

\*Corresponding Author: E-mail: stampfer@physik.rwth-aachen.de

[1] JARA-FIT and 2nd Institute of Physics, RWTH Aachen University, 52074 Aachen, Germany
[2] Peter Grünberg Institute (PGI-9), Forschungszentrum Jülich, 52425 Jülich, Germany
[3] National Institute for Materials Science, 1-1 Namiki, Tsukuba 305-0044, Japan

Graphene nanoribbons and constrictions are envisaged as fundamental components of future carbon-based nanoelectronic and spintronic devices. At nanoscale, electronic effects in these devices depend heavily on the dimensions of the active channel and the nature of edges. Hence, controlling both these parameters is crucial to understand the physics in such systems. This review is about the recent progress in the fabrication of graphene nanoribbons and constrictions in terms of low temperature quantum transport. In particular, recent advancements using encapsulated graphene allowing for quantized conductance and future experiments towards exploring spin effects in these devices are presented. The influence of charge carrier inhomogeneity and the important length scales which play a crucial role for transport in high quality samples are also discussed.

# 1. Introduction

Graphene - a two dimensional allotrope of carbon - has received a lot of attention in both fundamental research and industrial applications since its isolation in 2004. In contrast to many other materials, the band structure of graphene exhibits a linear dispersion at low energies. This unique electronic band structure confer with various extra-ordinary properties.[1-4] It is amongst the lightest, thinnest, most flexible and strongest materials in the world till date. It has a giant intrinsic carrier mobility [5], a huge thermal conductivity [6] and a high optical transparency [7] making it suitable for many ultra-thin, ultra-light opto-electronic applications [8].

In particular, in the field of microelectronics, where the miniaturization of devices is reaching its limits, graphene could be a particularly bright alternative to existing technologies. It is cross compatible, well patternable and has demonstrated a very high carrier mobility and near ballistic transport even at room temperature.[9, 10] Furthermore, its low effective mass and long spin diffusion lengths make it also interesting for spintronic applications.[11, 12] However, graphene lacks one crucial property when realizing many nanoelectronic device concepts - a band gap. Therefore, as such, it cannot be used in digital electronics which require high on-off current ratios. Hence, some of present day's research focuses on various approaches for achieving this goal. One rather simple and effective way to open a band gap is by means of electron confinement. For example, bottom-up synthesized atomically precise arm-chair graphene nanoribbons (GNRs) with intrinsic band gaps exceeding 2 eV have been reported.[13,14] It is important to note that the band gap depends sensitively on the edge structure of the nanoribbon and the GNR width. While for arm-chair GNRs, it is predicted that the band gap scales inversely with the width [15-18], zig-zag GNRs are expected to show conducting edge states. The latter are promising for exhibiting correlated low-dimensional magnetism.[19-24] All these makes it also interesting to study etched graphene nanoribbons (GNRs) or graphene nanoconstrictions (GNCs) (see Figure 1a and 1b). Important to note is that, as very narrow nanoribbons are needed to achieve large band gaps, edges show a considerable influence on the electronic transport of such devices. Furthermore, nanoribbons with rough edges together in conjunction with tiny islands of graphene (typically of 10-100 nm in size) form devices known as quantum dots which are predicted to exhibit long spin lifetimes making them suitable for future quantum computing.[25] Hence, understanding quantum transport through these devices is extremely useful for designing and optimizing the performance of electronic and spintronic quantum devices.

With the advent of improved fabrication technologies, ballistic transport has already been achieved at room temperature in high quality micron-scale encapsulated graphene devices.[9] More recently, also ballistic transport and size quantization effects in encapsulated graphene constrictions have been reported.[26] Such graphene quantum point contact devices could be exploited to construct valley filters and valley valves in analogy with spin filters and spin valves.[27] Moreover, with engineered electrostatic gating, these ballistic constrictions could also function as collimated point-like sources allowing to explore a wealth of electron optics experiments based on Veselago lensing, beam splitters, or waveguides.[28,29]

## 2. Scope of this review

The main focus of this review is on low-temperature quantum transport experiments on graphene nanoribbons and constrictions. In the first part, a general overview on transport experiments in GNRs and GNCs will be discussed beginning with devices on $SiO_2$, suspended devices, devices on hexagonal boron nitride (hBN) and finally graphene-hBN sandwich-based devices.

In the second part, we focus on the current state-of-the-art technology for fabricating high quality etched graphene nanoconstrictions by encapsulating it between two layers of hBN. Recent achievements supporting the high sample quality including high carrier mobility, ballistic transport and quantized conductance in etched devices are discussed. Emphasis is also laid on (i) discussing the limitations to observe high quality transport phenomena in the context of localized edge states and on (ii) modifying the local density of state and their direct physical manifestation in such devices. Thus, the aim of this article is to not provide an exhaustive review of available literature on GNRs but more to motivate the reader towards research in graphene nanoribbons for electronic applications by presenting the key contributions so far, highlighting the major challenges in this direction both in terms of device fabrication technology and the physics governing their electronic transport.

**3. Critical length scales and gate couplings in graphene nanodevices**

The properties of graphene nanostructures crucially depend on (i) the intimate environment, giving rise to charge carrier and strain inhomogeneities as well as on (ii) edge termination or edge roughness that likely give rise to localized trap states. In particular, the flatness of graphene or the suppression of nanometer-scale strain variations has been shown to be crucial for low doping and high carrier mobility [30, 31]. All these has important consequences on charge transport length scales in graphene as well as on gate couplings in realistic graphene nanodevices. Figure 2a shows, for example, how the net charge carrier density $n$ as function of back gate voltage $V_{bg}$ depends on the residual charge carrier $n_0$ leading to

$$n = \sqrt{(\alpha V_{bg})^2 + n_0^2}. \qquad (1)$$

Here, we assume that graphene is placed on a $d = 300$ nm thick SiO$_2$ layer deposited on a highly doped Si substrate that serves as gate electrode resulting in a gate lever arm $\alpha = \varepsilon_0 \varepsilon_r / d = 7.2 \times 10^{10}$ cm$^{-2}$ V$^{-1}$, where $\varepsilon_0$ is the dielectric constant and $\varepsilon_r = 4$. The residual charge carrier density reflects the presents of electron-hole puddles [32] near the charge neutrality point (CNP) most likely due to charge impurities in the near vicinity. In cleaner systems exhibiting low values of $n_0$ (see e.g. red trace in Figure 2a), there is less smearing of the charge carrier density around the CNP and a lower overall carrier density can be achieved. Figure 2b shows a similar plot where we additionally consider trap or localized states near the edges of a graphene nanostructure. In this case, we follow Ref. [26] and use

$$n = \frac{k_F^2}{\pi} = \alpha V_{bg} - n_T(V_{bg}), \tag{2}$$

where $k_F$ is the Fermi wave number and $n_T(V_{bg})$ is the accumulated trap state density. It is important to note that such trap states contribute to the total density of states (see inset of Figure 2b) and thus to the charging characteristics of the system but not to transport (see section 4.4 and Ref. [26] for a more detailed description about trap states). We see that as the trap state density increases the charge carrier density of the non-localized carriers smears out along the gate axis. In particular, we observe a gate dependent broadening of $n$ near the CNP. Taking into account both sources of possible "disorder", the important length scale for ballistic transport – the mean free path – is plotted as a function of applied gate voltage for different mobilities in Figure 2c. The mean free path is given by

$$l_m = \frac{\hbar \mu}{e} \sqrt{\pi n}, \tag{3}$$

where $e$ is the elementary charge, $\hbar$ the reduced Planck constant and $\mu$ the carrier mobility.[9] Figure 2c shows that for high carrier mobility devices e.g. $\mu = 300{,}000$ cm$^2$ V$^{-1}$ s$^{-1}$, it is possible to have ballistic transport over distances on the order of 1 μm at reasonable low carrier densities ($n > 9 \times 10^{10}$ cm$^{-2}$). For comparison, we also included the Fermi wavelength [26] given by

$$\lambda_F = \sqrt{4\pi/n} \, . \tag{4}$$

This length scale has to be significantly smaller than the mean free path ($\lambda_F \ll l_m$) for semi-classical ballistic transport. Interestingly, when reaching the regime where the mean free path exceeds the nanoribbon (or constriction) length ($l_m > l$) while the Fermi wavelength is on the order of the channel width ($\lambda_F \sim w$) we enter quantum ballistic transport where quantized conductance can be potentially observed (see below). This regime is highlighted by the shaded area for realistic device dimensions. For example, for a carrier mobility of 300,000 cm$^2$ V$^{-1}$ s$^{-1}$ realistic devices with a length of around 1.0 μm exhibit (quantum) ballistic transport while for $\mu$ = 150,000 cm$^2$ V$^{-1}$ s$^{-1}$ the required device size is around 0.5 μm. Thus, achieving a higher mobility relaxes the restriction on the size of the devices to observe ballistic transport. Figure 2d shows a similar plot as in Figure 2c but for different $n_0$ values all at $\mu$ = 150,000 cm$^2$ V$^{-1}$ s$^{-1}$.

It is important to note, that for (state-of-the-art) narrow (top-down) etched nanoribbons and constrictions (roughly $w < 200$ nm) the influences of the rough edges are playing a dominant and unfortunately limiting role [33-35]. For such devices localized states and statistical Coulomb blockade are opening a mobility gap making the mean free path in Figure 2c and 2d nearly meaningless. For the state-of-the-art etching technology, this puts a lower limit to nanoribbon and constriction width for observing quantum ballistic transport.

**4. Electronic transport in graphene nanoribbons and nanoconstrictions**

## 4.1. Graphene nanoribbons on SiO$_2$

The most commonly used substrate for the fabrication of GNRs and GNCs is a SiO$_2$ layer on highly doped Si. This is thanks to its wide availability, excellent dielectric properties of SiO$_2$ and the potential integration of graphene-based devices with the existing Si-based semiconductor technology. Most early experiments involved deposition of graphene on such Si/SiO$_2$ substrates either by direct exfoliation or by some other transfer method. GNRs or GNCs were then fabricated using a series of electron beam lithography and reactive ion etching (RIE) steps.[30-44] An example of a such fabricated graphene nanoribbon ($l = 2$ μm and $w = 80$ nm) is shown in Figure 3a. Other fabrication methods include unzipping of carbon nanotubes[45], bottom-up fabrication from molecules[46], growth along the edges of SiC steps[47], and epitaxial growth.[48] A comprehensive outline of various fabrication techniques and the relevant literature is mentioned in Refs. [49,50]. While bottom-up synthesized atomically precise GNRs are most promising for controlling the electronic properties, their maximum length sets some limitations for advanced transport studies and device integration. This makes, at present, top-down lithographically defined GNRs and GNCs, the most common approach for quantum transport structures.

Lateral confinement of electrons in lithographically patterned graphene nanoribbons is usually characterized by suppressed conductance particularly near the CNP leading to a so-called "transport gap" in back gate voltage characteristic.[37-40] Figure 3b shows an example of a conductance back gate characteristics highlighting the transport gap $\Delta V_{bg}$, centered around the CNP at $V_0$. In particular, statistical Coulomb blockade is observed in this region of suppressed conductance similar to the experiments based on quantum dots. Subsequently, transport in such disordered nanoribbons on SiO$_2$ was explained by treating the nanoribbon as a series of quantum dots of varying sizes that exhibit local resonances in the transport gap.[38] Moreover, it has also been successfully demonstrated, that the addition of single electrons to the nanoribbon can be monitored by a nearby single electron transistor with a combination of plunger and side gates. [51-54]. Associated with the region of suppressed conductance are two energy scales - one $\Delta E_F \propto \Delta V_{bg}$ that depends on the extent of the gap in the back gate voltage direction which is a measure of the disorder potential, the other, $E_g$ is the extent of bias voltage (in bias spectroscopy) which gives the charging energy of individual charged islands (see Figure 3c). The transport gap, i.e. the extent in back gate voltage $\Delta V_{bg}$ is a result of both the confinement of electrons due to etching and a bulk/edge disorder potential together resulting in the formation of charged islands and quantum dots. While the source-drain gap, $E_g$ is related to the size and charging energy of the associated quantum dots and therefore scales inversely with the width of the nanoribbon (narrower nanoribbons consist of smaller charge islands with larger charging energy).[37] This is in agreement with earlier theoretical calculations which suggest that the origin of energy gaps coming from both the confinement and disorder potential due to rough edges.[13] Further, scanning gate measurements also confirmed the existence of multiple quantum dots in such devices.[55]

Similar energy scaling was indeed also earlier observed by Han *et al*. [56] in nanoribbons with widths ranging from 10-100 nm and lengths of 1-2 µm (i.e., $w < l$). In another closely related study of aspect ratios in etched nanoconstrictions particularly in the regime of $w \sim l$ and also $w > l$ (smallest width = 50 nm, largest length = 1 µm) [57], shorter constrictions showed a much smaller transport gap. This is because shorter constrictions consisted of fewer charge islands and hence less charging events. Although the width dependent charging energy was nearly independent of the length, the minimum conductivity in the transport gap region itself was strongly length dependent with shorter constrictions having a much higher conductance level (0.1 - 1 e$^2$/h).

A temperature dependent study of the transport gap in nanoribbons with similar dimensions (20 nm < $w$ < 120 nm and 0.5 µm < $l$ < 2 µm) revealed that the transport through the nanoribbon is dictated by thermal activation of charge carriers at higher temperatures and variable range hopping at lower temperatures. Electric field-based transport measurements indicate that the transport gap in disordered GNRs is governed by hopping through localized states [58] or co-tunneling processes.[59]

Etched graphene nanoribbons were also synthesized using oxygen plasma reactive ion etching with a patterned hydrogen silsesquioxane HSQ layer as protective mask [60]. The transport properties were found to be heavily dependent on the proper removal of the HSQ layer using HF. In these devices having $w$ =30 nm width and $l$ = 1.7 µm channel length, first experimental evidence of sub band formation in 1D channels were claimed.[60] However, these devices are not ballistic which is due to a combination of high edge disorder (long channel length implies a stronger influence of edges) and also substrate-induced disorder. More recently, it has been shown that this edge disorder can be reduced by treating the etched graphene constrictions with a low concentration of hydrofluoric (HF) acid for a very short duration of 20 s.[61] This suppresses the transport gap significantly and shifts the CNP close to zero volts. However, the effective energy gap remains unaltered.

### 4.2. Suspended graphene nanoribbons

One straight forward approach to reduce substrate-induced disorder is to remove or to replace the SiO$_2$ substrate. The substrate could be either completely removed to form "suspended" nanoribbons or replaced by another more suitable material. The first significant improvement in device quality was observed by Bolotin *et al*. in micron-sized suspended and current annealed devices.[62] While the carrier mobility of devices before current annealing remained rather low (28,000 cm$^2$ V$^{−1}$ s$^{−1}$ at $n = 2\times10^{11}$ cm$^{-2}$), the mobility of current annealed devices was as high as 230,000 cm$^2$ V$^{−1}$ s$^{−1}$ at similar charge carrier density at a temperature of 4 K. This is in agreement with previous studies that current annealing removes residual dopants and other fabrication impurities thus resulting in a sharper resistance peaks at the CNP and, correspondingly, to a higher mobility and a lower charge inhomogeneity. Owing to the superior quality of such devices, many interesting phenomena such as ballistic transport,[63] snake states [29] and fractional quantum Hall effect [64] which were earlier limited by device quality are now observed.

Signatures of quantized conductance in graphene nanoconstrictions were first observed by Tombros *et al*. in devices suspended above the surface of a SiO$_2$/Si substrate.[65] These devices were current annealed which leads to a carrier mobility as high as 600,000 cm$^2$ V$^{−1}$ s$^{−1}$ at a charge carrier density of $5\times10^9$ cm$^{-2}$ at 77 K resulting in an electron mean free path of several hundred nanometers (200 - 450 nm).[66] Conductance steps were observed at intervals of around 2e$^2$/h suggesting that the valley degeneracy is lifted. The transition from quantized conductance at 0 T to the quantum Hall regime for magnetic fields above 60 mT was observed which confirms the high quality of these devices.[65] This work was further corroborated by calculations performed by S. Ihnatsenka and G. Kirczenow [67] which provide a theory for the observation of integer and fractional quantized conductance with a comparison between zig-zag and arm-chair edges. However, in devices produced by the same fabrication method and current annealing but with much smaller device width, Coulomb blockade was observed at 0 T similar to disordered nanoribbons fabricated on SiO$_2$. The Coulomb blockade becomes strongly suppressed beyond a relatively low magnetic field of nearly 2 T and shows a completely insulating state.[68] This shows that even for high quality samples, edge disorder most likely is inevitable and plays a key role in the determination of the observed physical phenomena. In similarly fabricated devices, formation of *p-n* junctions was also well studied. Characteristic Fabry-Perot interferences and a ballistic transport with a mean free path of nearly 2 µm were measured thus also paving the way for more advanced electron optics experiments in the future.[69,70] Quantized conductance was also observed in gate-defined bilayer graphene constrictions using multiple side and top gates. By using an electric field perpendicular to the layers, it is possible to open a bulk band gap and thus achieve quantum confinement without evoking edge disorder.[71]

Suspended devices, however, suffer from severe inherent limitations. Due to their fragileness they are prone to easy damage and the length of the devices is limited to a few micrometers. The suspended nanodevices described above are based on a current annealing step that actually also has been used to form the nanoconstrictions.[65] Crucially, this process does not allow for control over the dimensions of the devices. Also, the maximum applicable back gate voltage is in most cases limited to a few volts due to the straining of graphene. Furthermore, it is difficult to incorporate top gates and fabricate multi-terminal devices.

### 4.3. Graphene nanoribbons on hBN

Most graphene nanodevices fabricated from graphene that rest directly on $SiO_2$ suffer from inherent substrate induced disorder in the form of dangling bonds, substrate roughness, charge puddles [32,72] and surface phonons [73] which cause serious limitations to the mobility of charge carriers in such devices. Recently, hBN has received a great deal of attention as an alternative substrate for graphene owing to its atomically smooth surface that suppresses rippling in graphene, a lattice constant similar to that of graphene and planar structure that is expected to be free of any dangling bonds or surface charge traps.[74,75] Scanning tunneling microscopy studies also confirmed that graphene on hBN has significantly less pronounced electron-hole puddles as compared to SiO$_2$.[76]

However, incorporating a new substrate alone cannot improve the quality of nanostructure devices as reported by Bischoff *et al.*[33] It was seen that the micron sized graphene devices on hBN have significantly higher mobility (more than 45,000 cm$^2$ V$^{-1}$ s$^{-1}$ at n = 10$^{10}$ cm$^{-2}$) and lower disorder density (less than 10$^{10}$ cm$^{-2}$) than devices fabricated on SiO$_2$ [77]. However, the transport behavior of nanoribbons remained the same. In both cases, transport seems to be dominated by localized states and charge puddles similar to Ref. [38]. Further, nanoribbons of similar width etched from the same graphene flake showed strong variations in the quality and evolution of conductance. This is a strong indication that edge disorder plays an important role in the transport properties of reactive ion etched graphene nanostructures irrespective of the substrate. In order to reduce the influence of edges, the authors also fabricated ultra-short constrictions with a width of only 30 nm.[78] Analysis of transport gap revealed an interesting result - it is possible that the area over which the charge is localized can be almost 10 times larger than the constriction itself. This is allowed only if these localized states extend along the edges of the constriction into the graphene leads. In such a case, a small wave function overlap between the localized state in the edge and the delocalized state in the lead should allow electron tunneling. These findings further indicate the importance of the influence of edges on transport.

**4.4. Encapsulated graphene nanoribbons**

The most successful approach to completely isolate graphene from its surroundings for preventing contaminations is by encapsulating it within two flakes of hBN, thus forming a graphene-hBN "sandwich" structure. While the carrier mobility of graphene on SiO$_2$ was in the range of 10, 000 to max. 50,000 cm$^2$ V$^{-1}$ s$^{-1}$, the mobility of such sandwich structures reached the highest values of several millions [10], which is fully comparable to values obtained from suspended devices.[79,80] In contrast to suspended devices, sandwiched structures exhibit even room temperatures mobilities of up to 100,000 cm$^2$ V$^{-1}$ s$^{-1}$. [81] This technology therefore allows to achieve high quality devices and a good control over the geometry and size of the devices. Furthermore, it also allows to fabricate well-performing multi-terminal devices and to have an efficient gate coupling.[5,9,82]

In particular, dry transfer processes [5,83] for making hBN-graphene-hBN sandwiches in combination with metal (Al) hard mask and SF$_6$ or CHF$_3$ based reactive ion etching turns out to allow for well controlled constrictions with high electronic quality [26]. Examples of such fabricated hBN-graphene sandwich constrictions with several one-dimensional side contacts are shown as inset of Figure 4a and in Figure 5a.

For true quantum ballistic transport, we expect that the conductance can be described by the Landauer formula, given by [83,84],

$$G = \frac{4e^2 t_0}{h} \sum_{m=1}^{\infty} \theta\left(\frac{wk_F}{\pi} - m\right) \approx \frac{4e^2}{h} \frac{t_0 w k_F}{\pi}, \tag{5}$$

where $t_0$ is the average transmission per mode, $\theta(x)$ is the Heavy side step function and the factor 4 accounts for the spin and valley degeneracy. In short, we expect, in contrast to the diffusive transport, where $G \propto V_{bg}$ (see also red dashed lines in Figure 3b), a $G \propto k_F \propto \sqrt{V_{bg}}$ dependence. This square root dependence of the conductance on $V_{bg}$ can be, for example, well observed in Figure 4a. The red lines are square root fits to the conductance. By independently determining the gate coupling $\alpha$ (by the quantum Hall effect) and measuring the width $w \approx 280$ nm of the constriction (by scanning electron microscopy and atomic force microscopy) an average transmission $t_0 \approx 0.63$ can be extracted. The fact that the two red lines do not meet at $V_{bg} = 0$ V, i.e that they are offset on the back gate axis (see in particular close-up shown in Figure 4b) is explain by the presence of localized trap states not contributing to transport (see also black dashed lines in Figure 2b and Ref. [26]).

Figure 4b shows a close-up of Figure 4a highlighting kinks in the conductance trace (see arrows), which are due to the opening of individual conduction channels, i.e. modes $m$ (see Equation 5) when accessing sub-bands at higher energies. This is confirmed when studying the bias dependency, the Fourier transformation of $G(k_F)$ as well as the B-field dependency (for more details please see Ref. [26]). To further investigate the role of trap states at the edges and their influence on transport we fabricate encapsulated graphene constrictions with additional top gates covering the edges of the constriction. Top gates have already been demonstrated useful in controlling the electronic transport through structured graphene, for example, by forming a p-n junction that creates a potential barrier.[84,85] Figures 5a and 5c show atomic force microscope (AFM) images of a graphene constriction device with and without local top gates. For the latter, a layer of hBN (around 20-30 nm thick) is transferred on top of the contacted graphene constriction (300 nm width, 150 nm length) and acts as a top gate dielectric.[86] Two top gates, one on either side of the constriction are then patterned using standard e-beam lithography with a PMMA mask and subsequent Cr/Au evaporation. The top gates are designed to have a 30 nm overlap with the constriction area, also extending along the edges of the leads on either side of the constriction (Figures 5b, 5c and 5d). Because of their proximity, these top gates have quite a large capacitive coupling to the edges of the constriction. Therefore, this geometry allows to use the top gates to tune mostly the edge of the constriction and the area in its close vicinity, while the back gate tunes the entire device globally. Thus, this geometry provides additional knobs (two top gates) to tune the charge carrier density in the constriction. More importantly, in conjunction with the back gate, they can be used to study the filling up of localized edge states, i.e. trap states. Another difference in the geometry of the devices is the size of leads. While the size of the leads in Ref. [26] was only 1.5 – 2 µm other interference effects such e.g. Fabry-Perot oscillations which originate from the finite size of the leads were seen at very low temperatures to be super-imposed on the conductance kinks.[87,88] Hence, in the here presented devices with top gates, the size of the leads is increased to around 6 µm which ensures that the leads are significantly larger than the mean free path and the phase coherence length and that the transport through the leads is effectively diffusive.

Figure 6a shows a plot of the conductance $G$ measured through such a 300 nm wide constriction using a 4-probe geometry as a function of back gate $V_{bg}$ and top gate $V_{tg}$ voltages. As such there is no electrical contact between the two top gates, but in this particular configuration, they are short to each other externally. The white dotted line indicates the conductance minimum i.e. the CNP of each individual back gate trace in the plot measured at a particular top gate voltage. The slope of this dashed line is a measure of the relative lever arm $C_{bg}/C_{tg}$ i.e. the ratio of the capacitive coupling of the back and top gates to the constriction device. This map shows a non-constant relative lever arm in the voltage range under consideration. This is comparable to previous experiments on side gated HF-dipped graphene nanoribbons on $SiO_2$ (see also above).[61] In those devices, the non-constant lever arm has been attributed to edge modification due to their treatment with HF. Such a modification resulting in fluorine terminated graphene edges is also likely for the presented top gated devices since the sandwiches were etched using $SF_6$ during fabrication. Calculations also suggest that replacing –OH and –O by –F termination, leads to a significant reduction in density of states along the graphene edges resulting in gate dependent relative coupling.[61,89]

Figure 6b shows individual conductance traces (line cuts from Figure 6a) as a function of back gate voltage. For clarity, all traces are shifted by $V_0$ ($V_{tg}$) (back gate voltage at the CNP for every trace at a fixed value of $V_{tg}$) such that $\delta V_{bg} = V_{bg} - V_0$. As inset we show the quantity $t_0$ which has been extracted by fitting the conductance of individual back gate traces to a square root function (see Equation 1). Here we used a constant back gate lever arm of $7 \times 10^{10}\,\mathrm{cm^{-2}\,V^{-1}}$, which has been estimated by a plate capacitor model [90] and is in good agreement with values found on very similar devices (see table 1 in the suppl. material of Ref. [26]). Overall we conclude that the transmission through the constriction can be slightly tuned. This is most likely connected to the top gate dependent screening of the scattering at the rough edges in such constriction device, in good agreement with tight-binding calculations [26]. For further study in this direction, it would be highly informative to tune the two top gates independently from one another in both a symmetric and asymmetric manner to understand the influence of tuning multiple gates on the trap state density.[91,92]

Figure 6c shows a close up of the same traces as in Figure 6b but now constantly shifted in back gate voltage. Most importantly, we observe that the conductance steps/kinks on the order of 2 e²/h are nearly independent of $V_{tg}$ (see labels in Figure 6b) and thus stable signatures of quantized conductance.

Another interesting direction of work in high quality encapsulated graphene nanoconstrictions is the study of the evolution of the quantized conductance in the presence of a magnetic field. The evolution of the conductance steps in the presence of a perpendicular magnetic field was studied in detail in these devices, as shown in Ref. [26]. Landau fan measurements show a clear transition from size quantization at B = 0 T to Landau level quantization at higher magnetic fields similar to what was observed in suspended devices in Ref. [65]. The gate lever arm can be extracted from the slopes of the second derivative of the conductance in the Landau fan diagram. Moreover, bias spectroscopy measurements reveals "diamond" like features whose extend along the gate voltage axis correspond to the positions of two consecutive kinks. From these measurements, a sub-band spacing of around 14 meV has been extracted which is in agreement with a Fermi velocity in the order of $1.5 \times 10^6$ m/s [26].

## 5. Conclusion

Rapid technological developments over the last decade have immensely contributed to improving the quality of graphene nanoribbons and constrictions. However, controlling and suppressing the edge disorder will play an important role in understanding electronic transport through these devices. Fabricating multiple side and top gates allows to obtain more information such as the coupling and density of localized edge states. Since the exact nature of edges is so far not known, it would be ideal to combine various experimental techniques such as quantum transport at low temperatures, Raman spectroscopy, and scanning tunneling microscopy on the very same device. Other alternative forms of manipulating the edges such as edge functionalization/passivation could also be explored.[93,94] Finally, it is also interesting to look at the quality of graphene nanoribbons on other two dimensional substrates besides hBN such as $MoS_2$, $WS_2$ and others.[82] Such materials also offer an opportunity to study proximity effects in graphene which can open a spin orbit gap that can be utilized for spintronics.[95]

**Acknowledgements** Support by the Helmholtz Nanoelectronic Facility (HNF), the EU ITN SPINOGRAPH and the DFG (SPP-1459) is gratefully acknowledged. Growth of hexagonal boron nitride crystals was supported by the Elemental Strategy Initiative conducted by the MEXT, Japan and JSPS KAKENHI Grant Numbers JP26248061, JP15K21722 and JP25106006.

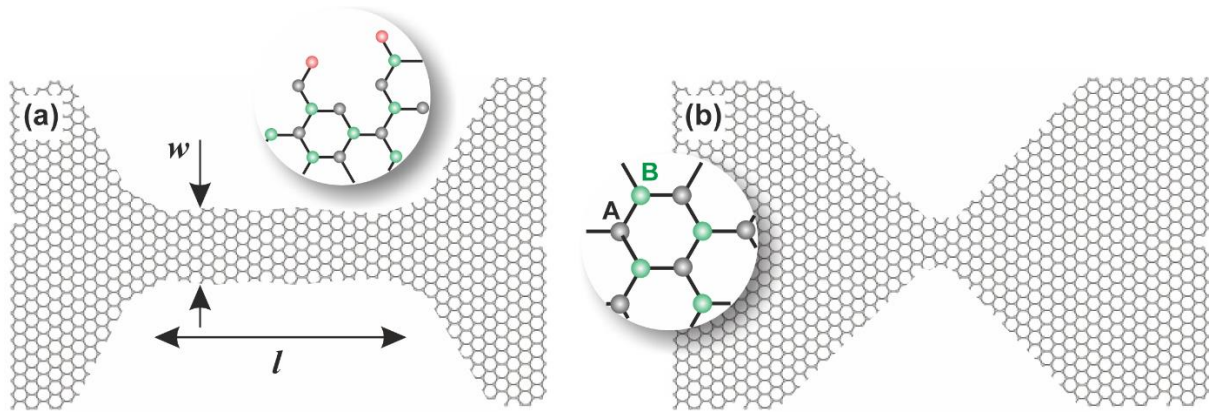

**Figure 1.** Schematic illustration of a graphene nanoribbon (a) and nanoconstriction (b) of length $l$ and width $w$. Inset in (a) indicates rough edges and dangling bonds (red dots) along the edge of the nanoribbon. Inset in (b) shows the hexagonal lattice structure of graphene with carbon atoms in the two sub-lattices A (grey dots) and B (green dots).

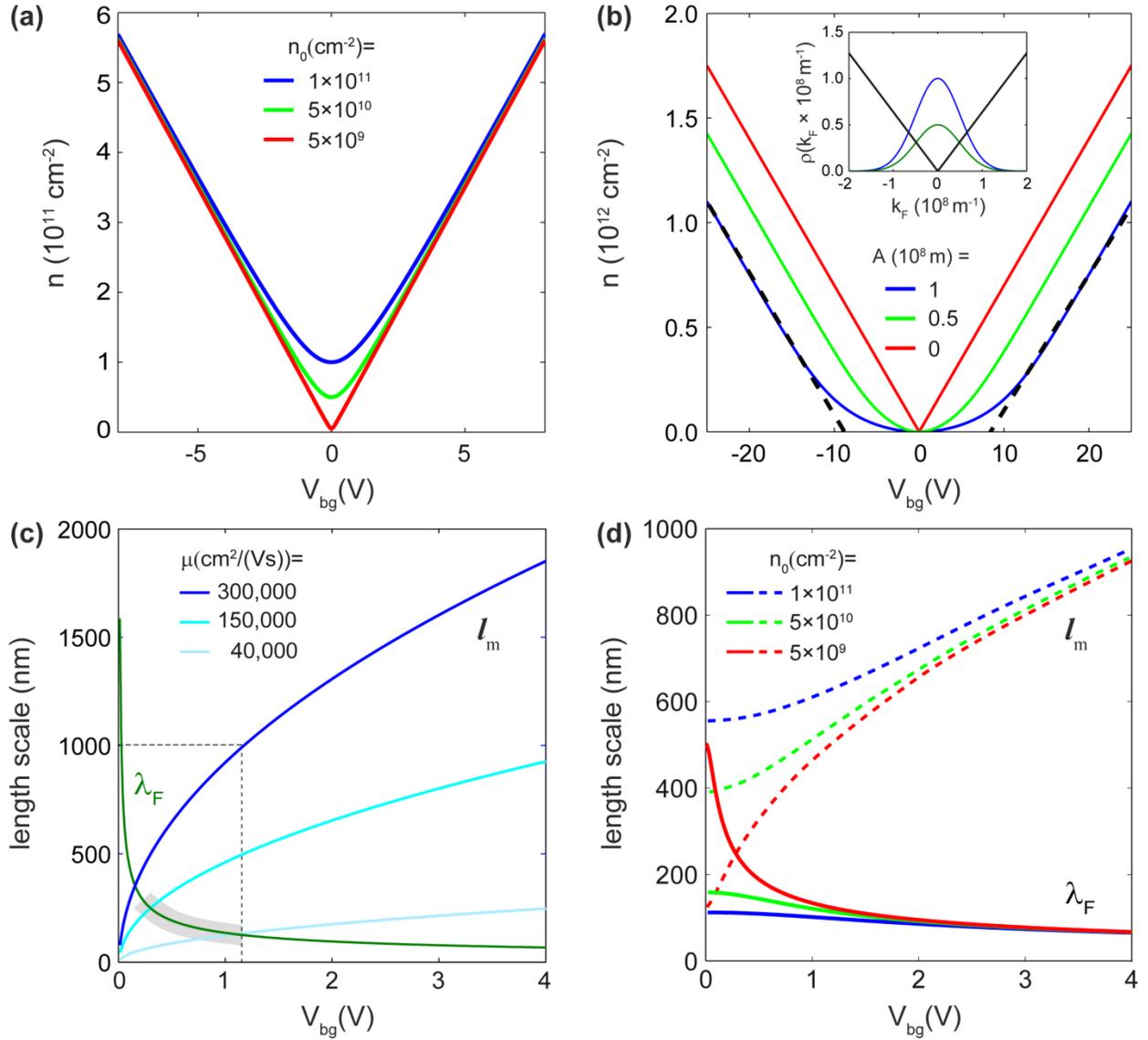

**Figure 2.** (a) Charge carrier density $n$ as a function of the back gate voltage $V_{bg}$ for different residual charge carrier density $n_o$. (b) $n$ as a function of $V_{bg}$ for different trap state carrier density ($n_T$). The black dotted line indicates the deviation from the linear behavior of the charge carrier density around the CNP. Inset shows the density of trap states (blue and green traces correspond to Gaussian distribution functions with different amplitudes A). The solid black trace corresponds to that of graphene. (c) Fermi wavelength $\lambda_F$ (green trace) and mean free path $l_m$ (blue traces) as a function of $V_{bg}$ for different mobilities with $n_o = 5 \times 10^9$ cm$^{-2}$. (d) Similar to (c) with same mobility $\mu = 150{,}000$ cm$^2$ V$^{-1}$ s$^{-1}$ but variable $n_o$.

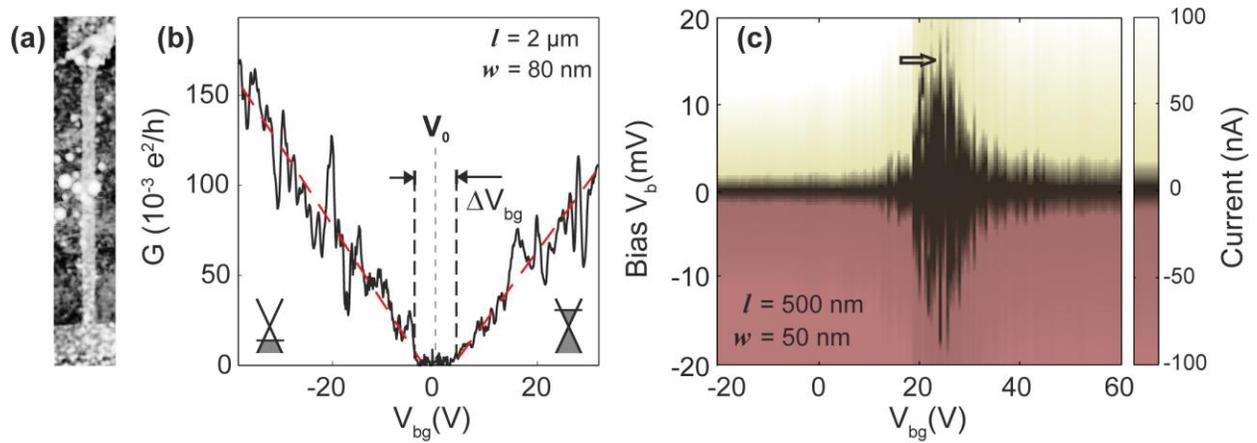

**Figure 3** (a) Scanning electron micrograph of graphene nanoribbon with $l = 2$ μm and $w = 80$ nm. (b) Two-terminal back gate characteristics of the graphene nanoribbon shown in panel (a). The hole and electron transport are highlights by the insets and red dashed lines mark the overall linear increase of conductance. (c) Color plot of the source-drain current as a function of the back gate voltage $V_{bg}$ and the bias $V_b$ for a constriction with w = 50 nm and l = 500 nm. The arrow indicates the region of suppressed current which comprises of Coulomb diamonds. Figure (c) is reprinted from Appl. Phys. Lett. **98**, 032109 (2011) with the permission of AIP Publishing.

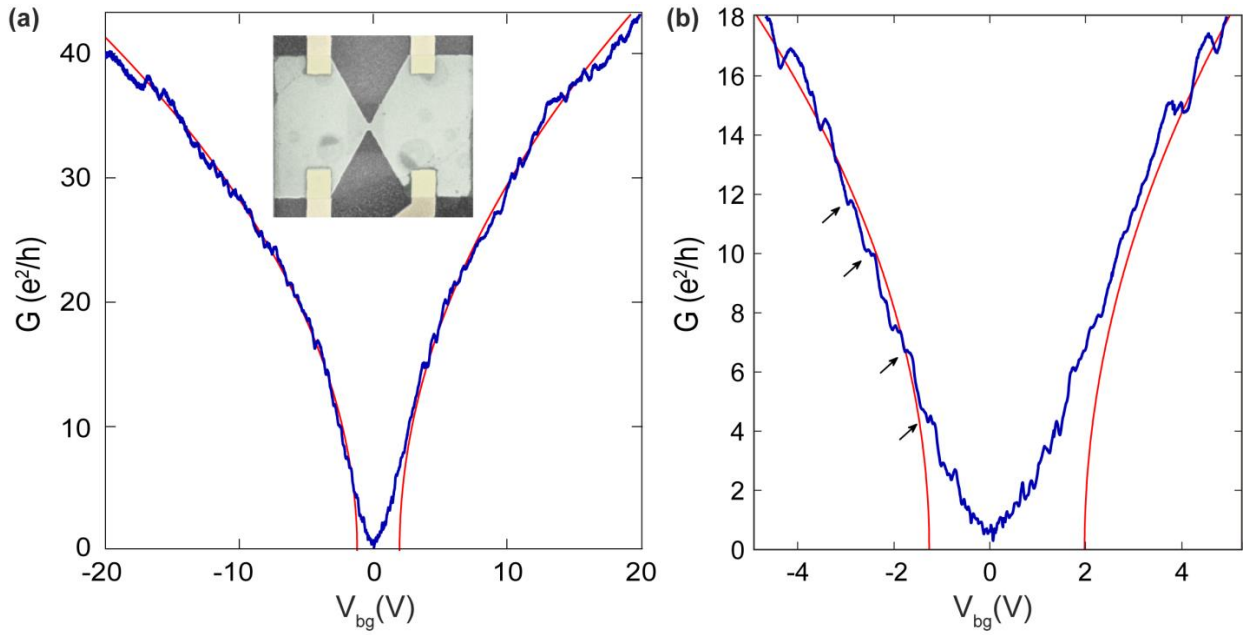

**Figure 4.** (a) Conductance trace of a 280 nm wide encapsulated graphene nanoconstriction (inset) as a function of the back gate voltage (blue trace) taken at T = 4 K. (b) Close-up of the trace in (a) around the CNP. Conductance kinks in steps of nearly $2e^2/h$ as denoted by the black arrows are observed. Deviation of $G \propto \sqrt{V_g}$ relation obtained from Landauer theory (red trace) is visible around the CNP.

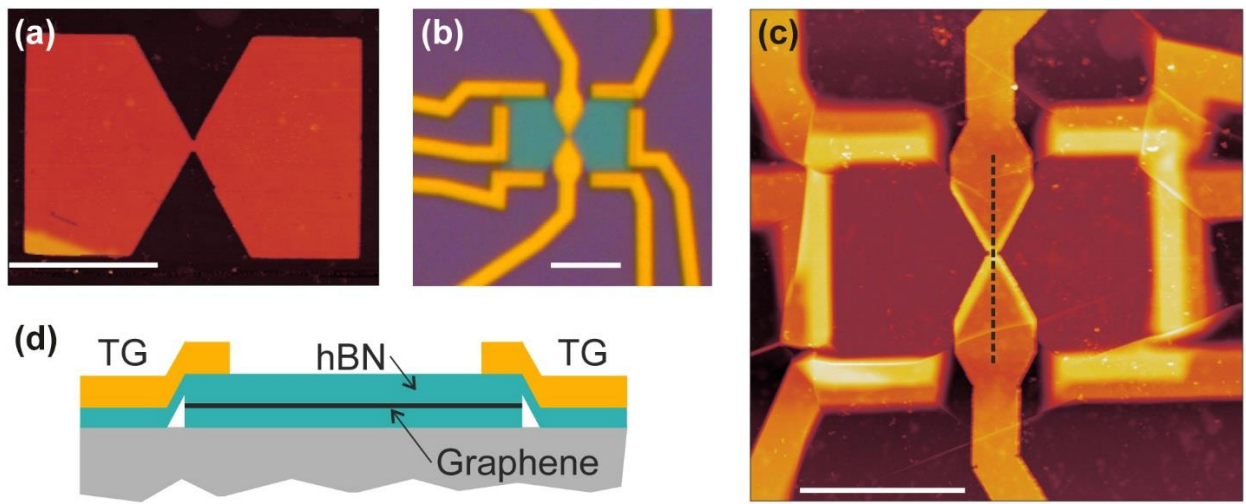

**Figure 5.** (a) Atomic force microscopy (AFM) image of an etched 300 nm wide hBN encapsulated graphene constriction. (b) Optical microscope image of the final device. The sandwich is blue, the gold contacts are yellow and the SiO$_2$ substrate is purple in color. The device consists of six side contacts- three on each lead and two additional top gates- one on either side of the constriction (marked by yellow arrows in panel (c)). (c) AFM image of the device in panel (b). Scale bars correspond to 4 μm. (d) Schematic cross-sectional illustration of the device along the dashed line in (c) highlighting the top gate geometry.

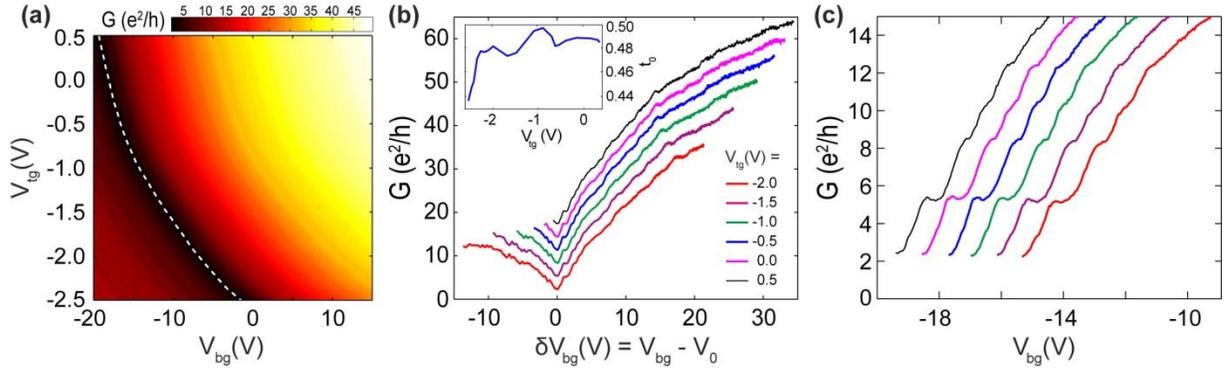

**Figure 6.** (a) Color-scale plot of the conductance $G$ of a 300 nm wide encapsulated nano-constriction as a function of the back gate $V_{bg}$ and the top gate voltage $V_{tg}$ (The two top gates are swept simultaneously). White dashed line indicates the CNP of each conductance trace in back gate measured at a fixed value of the top gate voltage and is a measure of the relative lever arm. (b) Horizontal line-cuts of the 2D plot in panel (a). Individual curves represent back gate traces at different top gate voltages with the CNP shifted to zero. The curves are offset vertically by 4 $e^2/h$ for clarity. Inset shows the transmission coefficient calculated from individual back gate traces as a function of $V_{tg}$. (c) The same traces as shown in the main figure (b) but shifted horizontally due to the applied top gate voltage. The black trace is at a fixed value of $V_{tg} = 0.5$ V and all the other traces to its right are shifted by 0.85 V relative to the preceding trace.